# Hot embossing of Au- and Pb- based alloys for X-ray grating fabrication

Lucia Romano[a]

Paul Scherrer Institut, 5232 Villigen PSI, Switzerland;
Institute for Biomedical Engineering, University and ETH Zürich, 8092 Zürich, Switzerland;
Department of Physics and CNR-IMM- University of Catania, 64 via S. Sofia, Catania, Italy

Joan Vila-Comamala

Paul Scherrer Institut, 5232 Villigen PSI, Switzerland;
Institute for Biomedical Engineering, University and ETH Zürich, 8092 Zürich, Switzerland

Helmut Schift

Paul Scherrer Institut, 5232 Villigen PSI, Switzerland

Marco Stampanoni

Paul Scherrer Institut, 5232 Villigen PSI, Switzerland;
Institute for Biomedical Engineering, University and ETH Zürich, 8092 Zürich, Switzerland

Konstantins Jefimovs

Paul Scherrer Institut, 5232 Villigen PSI, Switzerland;
Institute for Biomedical Engineering, University and ETH Zürich, 8092 Zürich, Switzerland

[a] Electronic mail: lucia.romano@psi.ch

**Abstract:**

Grating-based X-ray phase-contrast interferometry has a high application impact in material science and medicine for imaging of weakly absorbing (low Z) materials and soft tissues. For the absorbing gratings, casting of highly x-ray absorbing metals, such as Au and Pb alloys, has proven to be a viable way to generate large area periodic high aspect ratio microstructures. In this paper, we review the grating fabrication strategy with a special focus on a novel approach of casting low temperature melting alloys (Au-Sn and Pb-based alloy) into Si grating templates using hot embossing. The process, similar to nanoimprint lithography, requires particular adjusting efforts of process parameters as a function of the metal alloy and the grating feature size. The transition between solid and liquid state depends on the alloy phase diagram, the applied pressure can damage the high aspect ratio Si lamellas and the microstructure of the solid metal can affect the grating structure. We demonstrate that metal casting by hot embossing can be used to fabricate gratings on large area (up to 70×70 mm$^2$) with aspect ratio up to 50:1 and pitch in the range of 1-20 μm.

## I. INTRODUCTION

Grating-based X-ray phase-contrast Interferometry (GI) enables phase contrast imaging with a much higher image contrast of weakly absorbing materials (such as soft tissues) than conventional absorption-based X-ray radiography imaging[1] with relevant impact in material science and medical applications. The main challenge is the fabrication of the absorption gratings[2], which are metal periodic microstructures with high aspect ratio. There is a need to fabricate micro-gratings with i) high aspect ratio (in the range of 100:1, structural width in the micrometer range); ii) large area (mammography, e.g., requires for a field of view of 200 × 200 mm$^2$) and iii) good uniformity (no distortions and change in the period and height over the whole grating area). Absorption gratings are usually fabricated by metal electroplating (typically of Au, which is one of the most efficient absorbing materials for X-rays), into high aspect ratio grating templates produced by deep X-ray lithography (also called LIGA)[3] or deep Si etching[2]. Microcasting for X-ray absorption gratings has been developed by using molten Bi (melting temperature 271 °C) via capillary action and surface tension[4-6]. However, the low density (9.78 g/cm$^3$, atomic number 83) of Bi requires much higher (factor of 1.7 at 30 keV) aspect ratio structures to get an absorption comparable to that of Au (density 19.32

g/cm$^3$, atomic number 79). In a previous paper[7] we demonstrated that metal microstructured optical elements for GI can be fabricated by using an alternative approach of Au-Sn microcasting into Si templates, which has the advantage of fast processing with a competitive cost and X-ray absorption only 20% less than pure Au (at X-ray energy of 30 keV, see table 1). Metal casting can potentially be realized with any metal whose melting temperature is compatible with the available hot embossing tool and grating substrates. However, apart from cost issues and availability of suitable alloys for specific applications, differences are expected in terms of melting behavior, wetting, and morphology. In this work, we therefore extended the experiments to Pb (density 11.34 g/cm$^3$, atomic number 82)-rich compounds in comparison to Au-Sn alloy and we investigated the limits of the process in terms of grating stability, aspect ratio and uniformity. We analyzed in details the material properties as a function of the casting process in Si gratings for both classes of alloys: Au-Sn and Pb-based alloys. Three main material properties turned out to be relevant for the realization of metal gratings by hot embossing: i) wettability of the grating surface with respect to the liquid metal used for casting; ii) phase diagram of the metal alloy in order to properly set up the temperature range and process conditions; iii) thermal stress and fracture behavior due to the mismatch in thermal expansion coefficients between the poly-crystalline metal and the crystalline Si. The three aspects are presented and discussed with few examples that helped to generalize the required material properties and grating design for metal casting fabrication. X-ray preliminary characterization has been performed in order to investigate the uniformity of the metal filling on the grating area and the quality of the casted gratings in the interferometric set-up.

## II. EXPERIMENTAL

Metal microgratings were fabricated according to the work flow described in Figure 1. The pattern was designed by photolithography on Si <100> substrates (diameter of 100 mm) and the silicon etching was performed by deep reactive ion etching (Bosch process)[8] or metal assisted chemical etching (MacEtch)[9,10], the process details are reported in refs.[8-10]. The pitch of the grating was in the range of 1-20 µm and the duty cycle was 0.5, the grating area was 70 × 70 mm$^2$. A conformal layer of Ir (20-30 nm Ir on 10 nm $Al_2O_3$) was later deposited by Atomic Layer Deposition (Picosun R-200 Ad.) to improve the wettability of the Si template with respect to liquid metal alloy[7]. The metal casting was performed in a Jenoptik HEX 03 hot embossing tool, in vacuum (pressure in the range 100-500 Pa), by pressing a metal foil with same size as the grating in contact with the grating surface at the melting temperature of the metal foil. A 1 mm thick sheet of silicone rubber (PDMS, i.e. polydimethylsiloxane) was used as a cushion layer for pressure equilibration and to smoothen out any kind of unevenness, e.g. caused by substrate bow and warp or even dust particles. A polished Si chip with thickness of 500 µm was used as a flat surface to apply the pressure on the metal foil. The pressure was varied from 1 MPa to 12 MPa. The temperature was controlled with a precision of 2 °C. Further details of the casting process are reported elsewhere.[7] Test samples with the size of 20 × 20 mm$^2$ were cut from the original wafer. Optimized process was realized on full wafer with grating area of 70 × 70 mm$^2$ (see Fig.1). We used metal foils of eutectic Au-Sn (80 w% Au – 20 w% Sn with melting temperature of 280 °C) alloy from Ametek; Pb-In (95 w% Pb – 5 w% In, *liquidus* point 314°C, *solidus* 292 °C) and Pb-Sn-Ag (92.5 w% Pb – 5 w% Sn – 2.5 w% Ag, *liquidus* point 296 °C, *solidus* 287 °C) compounds from Hi-Rel Alloys. For minimizing the excess of material, the thickness of the metal foil has to match the cavity volume of the grating. For Au-Sn foils it is possible to choose the thickness in the range of 25 - 50 µm, while for Pb-based alloys we used foils with thickness of 100 µm, since thinner foils are usually not commercially available in size of 70 × 70 mm$^2$.

Scanning Electron Microscopy (SEM) Zeiss Supra VP55 was used to characterize the casted gratings in cross section. The X-ray performance of Au-Sn casted grating with a pitch of 4.8 µm was investigated by setting up a laboratory X-ray grating interferometry system using a Hamamatsu L10101 X-ray microsource (35 kV voltage and 0.2 mA electron current) and scintillator-fiber-optic coupled CCD X-ray detector (1024x1024 pixels of 13 µm).

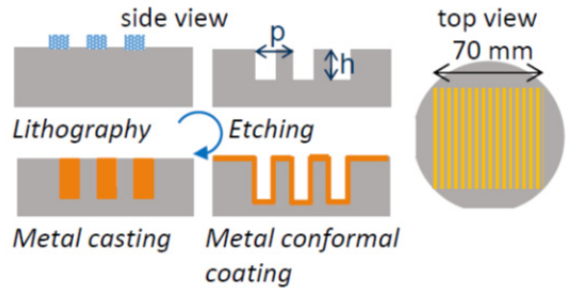

FIG. 1. (Color online) Schematic of the different stages of the fabrication sequence (not to scale): resist pattern by lithography, etched trenches, trenches/grooves with metal conformal coating and completely filled trenches with zero excess material. The top view shows the grating area (70x70 mm$^2$) on a silicon wafer with diameter of 100 mm.

## III. RESULTS AND DISCUSSION

### A.  *Metal alloy properties*

The wetting properties of the liquid metal both on top and on the trench wall surface were extremely critical for uniformly filling the Si grooves[7,11]. This was demonstrated by a series of experiments with different metal alloys and metal coating of the Si gratings. When the liquid metal did not uniformly wet the Si surface the metal was unequally distributed over the grooves, filling only some of them. According to the Young-Dupré equation[12], good wetting (i.e., a contact angle of few degrees) of a liquid metal on a solid substrate can be observed if the adhesion energy is close to the cohesion energy of the liquid. This condition is fulfilled for liquid metals on solid metals regardless of the miscibility between the liquid and the solid, because the interfacial bond is metallic[12]. The wetting behavior of the Si template with respect to liquid Au-Sn alloy was realized by a conformal coating of the Si grating with a metal film. In the reported experiments, we used Ir films deposed by ALD. The wetting properties of Au-Sn alloy on the Si grating were studied in a previous publication[7]. The Au-Sn alloy (80 w% Au – 20 w% Sn) has an eutectic transition at the temperature of 280 °C. This means that the solid metal foil contains the typical lamellar microstructure of eutectic alloy, the grain size for Au-Sn are usually few hundred nm or even smaller. The solid-liquid transition occurs at the eutectic temperature, ensuring a fast and sharp phase change at the melting temperature. The strategy to obtain metal casting with eutectic alloys was to apply the pressure during the solid-liquid transition in order to force the liquid metal to flow into the Si grooves and to reduce the flow on the grating top surface[7]. The overall process time was about 30 min, including heating and cooling. An example of this approach is reported in a previous publication[7].

For Pb-In alloy, a wetting layer can be created during the thermal process of hot embossing without preliminary ALD coating. This property is due to the presence of In in the compound and because In is a good dopant of Si that forms a chemically stable bonding on the Si surface. The Si grating is dipped in a water diluted HF solution (10 w%) for 10 min immediately prior the hot embossing. The phase diagram of Pb-In alloy with composition of 95 w% Pb and 5 w% In indicated the *solidus* line at 292 °C and the *liquidus* line at 314 °C[13]. With a sufficient slow heating (10°C/min) and a long process time (600 s under pressure in the temperature range of 290 – 330 °C) in the hot-embossing system, the separation of the In-rich phase, according to the phase diagram[13], with respect to the Pb-rich phase can be observed. Figure 2 (a) shows the In-rich nanoparticles formed on the Si surface. The liquid Pb-rich compound does not wet Si as indicated by the contact angle (>90°) in Figure 2 (b). However, if the liquid Pb-alloy flows on the In coating, the system can realize a wetting contact, as indicated in Figure 2 (c). Since the alloy has a very small percentage of In (5 w%), the In segregation at the Si surface at the beginning of the melting does not significantly affect the final solid composition of the alloy. These results indicated that a sufficient slow heating and a long process time are the parameters to play in order to optimize the metal filling without the requirements of any additional ALD wetting layer.

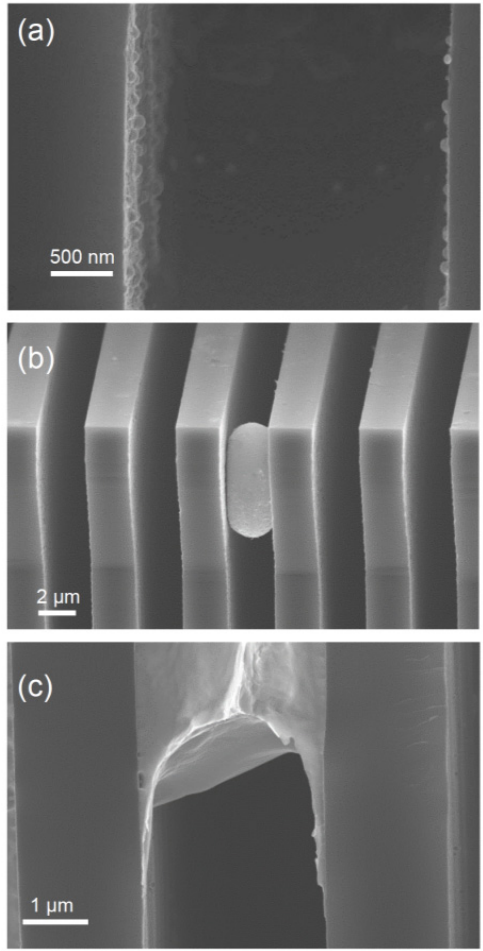

FIG. 2. Pb-In alloy can be separated during melting by a slow heating (10°C/min). The micrographs are high magnification SEM cross sections of grating trenches where: (a) the In-rich nanoparticles are formed on the HF-cleaned Si surface; (b) the liquid Pb does not wet Si as indicated by the contact angle (>90°); (c) the liquid Pb-alloy flows on the segregated In coating, which serves as a wetting layer for the Pb. Therefore, the system can realize a wetting contact without significant variation of the composition or even depletion of the alloy. The grating has pitch of 4.8 μm and was realized by Bosch etch.

An example of optimized process is reported in Figure 3. The slow heating is realized with a temperature gradient between the top and the bottom plate of the hot embossing tool, the used temperature ramps are showed in Figure 3 (a). The higher temperature of the bottom plate with respect to the top plate promotes the melting of the metal foil from the surface in contact with the Si grating. Figure 3 (b) shows the Si grating filled with Pb-In alloy, the Si grating was realized by MacEtch, the pitch is 6 µm and the trench depth is 60 µm. The metal casting was realized by applying a pressure of 5 MPa for 600 s.

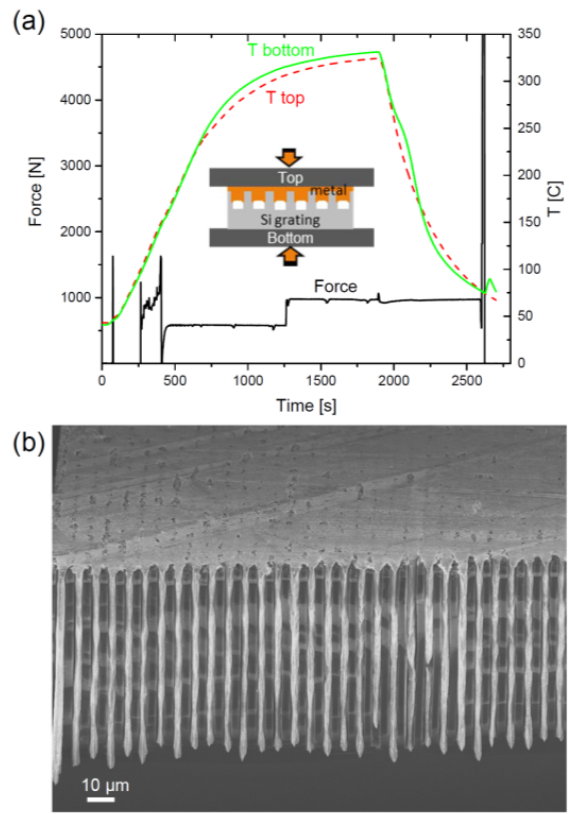

FIG. 3. (Color online) (a) Plot of the hot plate temperature, the bottom plate is in contact with the Si template, the top one with the metal foil as indicated in the insert, the applied force is also reported, the pressure is about 5 MPa; (b) Pb-In casted grating, the grating area was 20 × 20 mm$^2$, the pitch is 6 µm and the depth 60 µm, the grating was realized by MacEtch.

For the ternary Pb-Sn-Ag alloy, the realization of small pitch gratings is affected by the presence of large micro-grains in the foil. During the heating, when the temperature of the process approaches the *solidus* curve in the phase diagram, a liquid containing solid microparticles is formed. Once the *liquidus* curve is reached the full melting of the metal alloy is realized. The main difference with the eutectic (Au-Sn) system is that the melting is realized in a range of temperatures instead of a single temperature. If the metal is squeezed by hot embossing during the phase between *solidus* and *liquidus*, the solid microparticles can damage the Si template depending on the relative size. An example of this effect is reported in Figure 4. The metal foil was Pb-Sn-Ag alloy (92.5 w% Pb – 5 w% Sn – 2.5 w% Ag, liquidus point 296 °C, solidus 287 °C), the foil was embossed by applying a pressure of 5 MPa in the temperature range of 285 - 330 °C for 500 s. The microparticles have size in the range of few micrometers and they are probably Ag segregation since Ag is the metal with the higher melting temperature. Some microparticles were squeezed inside the Si grating with pitch of 4.8 μm during hot embossing and they deformed the grating lines indicating that this kind of metal alloy is not suitable for the realization of grating with trench size smaller than 3 μm.

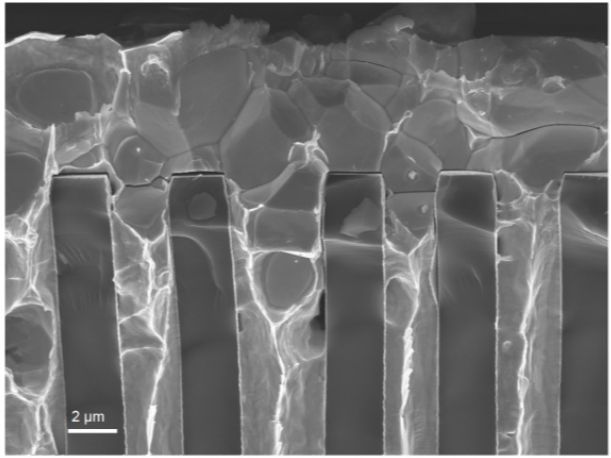

FIG. 4. Pb-Sn-Ag alloy (92.5 w% Pb – 5 w% Sn – 2.5 w% Ag) casted in Si grating with pitch size of 4.8 µm realized by Bosch etch and conformally coated by 30 nm Ir. Hot embossing pressure 5 MPa at a maximum temperature of 330 °C.

The phase diagram of the metal alloy is used to determine the melting temperature range for the hot embossing experiments. For non-eutectic compounds, the microstructure of the metal foil can be extremely relevant for the realization of the metal gratings by metal casting in Si templates.

## B. Thermal stress

The metals and the Si have different thermal expansion coefficient (Au-Sn[14]: $16 \cdot 10^{-6}$/C; Pb: $28 \cdot 10^{-6}$/C; Si: $3 \cdot 10^{-6}$/C). During the cooling down, the volume shrinkage of the casted metal lines can be easily observed in the large pitch gratings of Au-Sn (pitch 20 µm) and Pb-In (pitch 4.8 µm) reported in Figure 5.

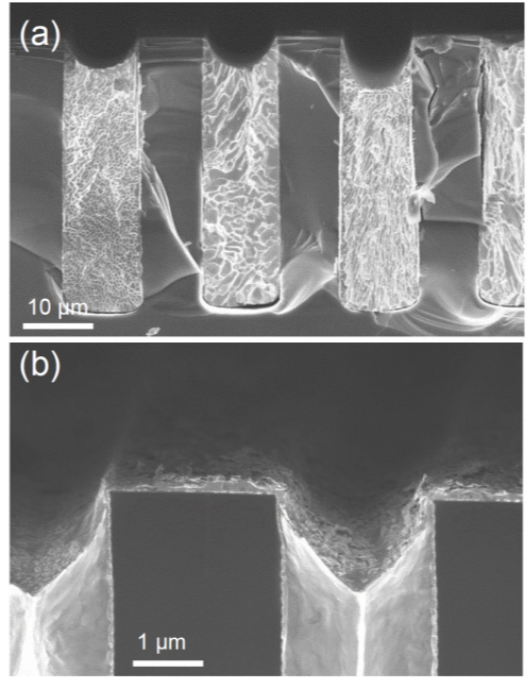

FIG. 5. Au-Sn (a) casted grating with pitch size of 20 μm; Pb-In (b) casted grating with pitch size of 4.8 μm. In both gratings the large grooves are formed on top of the metal lines because of the metal shrinkage during the cooling down.

The stress released at the interface with Si can cause wafer bowing and cracking. In particular, the brittle nature of crystalline Si with respect to polycrystalline metals leads to the Si fracture[15]. The stress release causes the cracking of the substrate when the metal is very stiff, which is the case for the Au-Sn alloy (young modulus 68 GPa[14]). This is shown in Figure 6, the Si grating is cracked along the (111) and (110) planes and a part of the grating is pulled off with a visible tilt. The pulling off is visible for several millimeters in optical microscope (not reported) by inspecting the grating surface immediately after the casting process, so the cracks are not caused by cleaving of the substrates for the SEM cross-section preparation. The cracks are regularly present with a period of few hundreds micrometers, as indicated in Figure 6 (c). As a consequence, the Si cracking affects the grating performance in the X-ray interferometry. In order to prevent the Si cracking, we performed a very slow cooling step in order to relax the thermal stress. The cooling rate is reported in Figure 6 (a), the fast cooling (black continuous line) produced the cracks and the piling up that are reported in Figures 6 (b-c), while the slow cooling (red dotted line) led to intact Si (Fig.6d-e). With slow cooling, the stress was released and only produced the metal delamination in some lines (Fig. 6d-e). Delamination does not significantly affect the X-ray performance of the grating since there are no distortions to the surrounding grating lines. Effectively, such a metal delamination has the same effect of local linewidth change for this particular line.

Since the Pb-rich alloys are less stiff (Pb young modulus: 16 GPa), the Si cracking is not observed in this case.

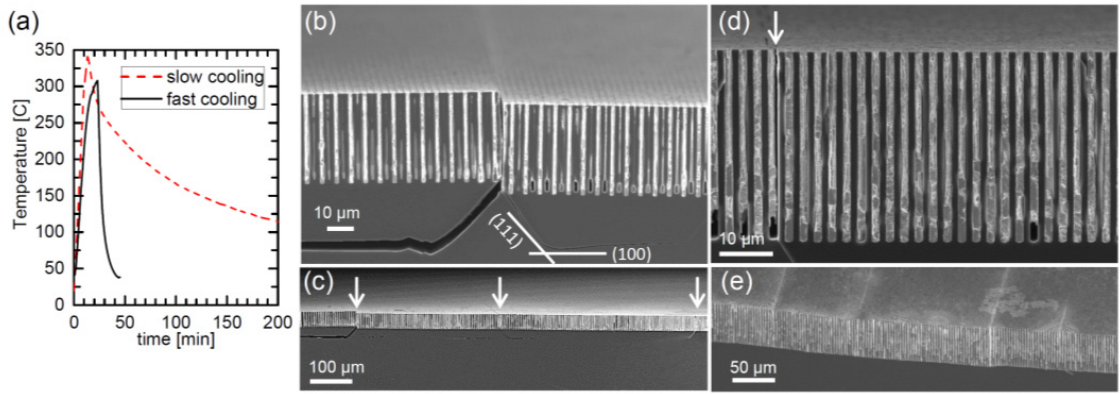

FIG. 6. (Color online) Temperature (a) vs time used for fast and slow cooling experiments. Au-Sn casted grating (b-c) obtained with fast cooling rate. The direction of the Si planes (111) and (110) are plotted as a reference in (b), the arrows in (c) indicate the cracking of Si substrate along the (111) and (100) directions and grating pulling off due to the release of the thermal stress. Au-Sn casted grating (d-e) obtained with the slow cooling rate, the arrow points to delamination, the low magnification image (e) shows no cracks in Si but only delamination.

## C. High aspect ratio structures

The pressure of the in-printing process is a relevant parameter to ensure the complete filling of the cavities. The process is realized in vacuum (chamber pressure in the range 100-500 Pa) to avoid the presence of air bubbles inside the grating lines. However, some cavities are formed inside the metal lamellas because of irregularities in the metal foils or metal shrinkage during the cooling phase. Figure 7 shows a comparison of Pb-In casting realized with embossing pressure of 0.4 MPa (Fig. 7a) and 1.5 MPa (Fig. 7b) into Si grating with pitch size of 4.8 μm.

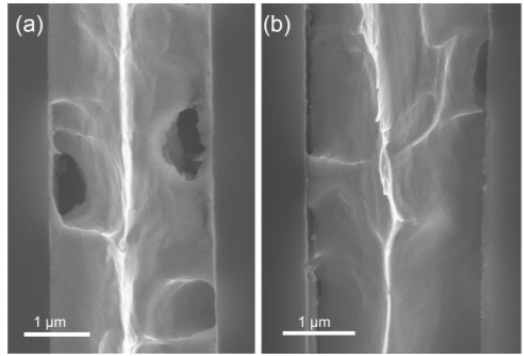

FIG. 7. Pb-In casting realized with embossing pressure of (a) 0.4 MPa and (b) 1.5 MPa into Si grating with pitch size of 4.8 μm.

High pressure is definitely required to realize metal casting in high aspect ratio gratings, by which the liquid metal is predominantly pushed into the grooves until the voids are filled and the foil volume is completely displaced. This seems to be valid both for eutectic, binary and ternary metals, independently of their different melting behavior. The achieved aspect ratio for all metals is about 50:1. Some examples are reported in Figures 8. Figure 8 (a) shows a Si grating with pitch of 1 μm and trench depth of 30 μm realized by Bosch process, coated with Ir layer by ALD and casted with Pb-In at a pressure of 5 MPa. In this case, the ALD process was performed to ensure the best uniformity of wetting layer, being the structure with very high aspect ratio and very small pitch (1 µm). Due to the high aspect ratio, the Si lines deformed by bending on one side, the applied pressure was not supported without deformation of the grating. Figure 8 (b) reports the same effect for a grating with pitch of 6 µm and height of 140 µm realized by MacEtch and casted with Pb-In at 5 MPa. In both cases, a good metal filling was achieved, as the metal can be seen all the way down to the bottom of the Si grooves. The two examples have the same aspect ratio of 50:1 with different pitch size, indicating that the smaller pitch size does not limit the aspect ratio of the casting. However, since the filling is complete, which can only occur with an intact grating, i.e. with vertical lamellas, the bending is probably caused by a hard contact between the upper stamper and the grating or an increase of the shearing force while the foil has only a fraction of its initial thickness (Figure 8 indicates that there are residues of < 1 µm). The shearing enables a further compression and thus a bending, and possibly a squeezing out of the casted material from the trenches. Therefore, by using thicker foils it is not unlikely that the aspect ratio can be further enhanced.

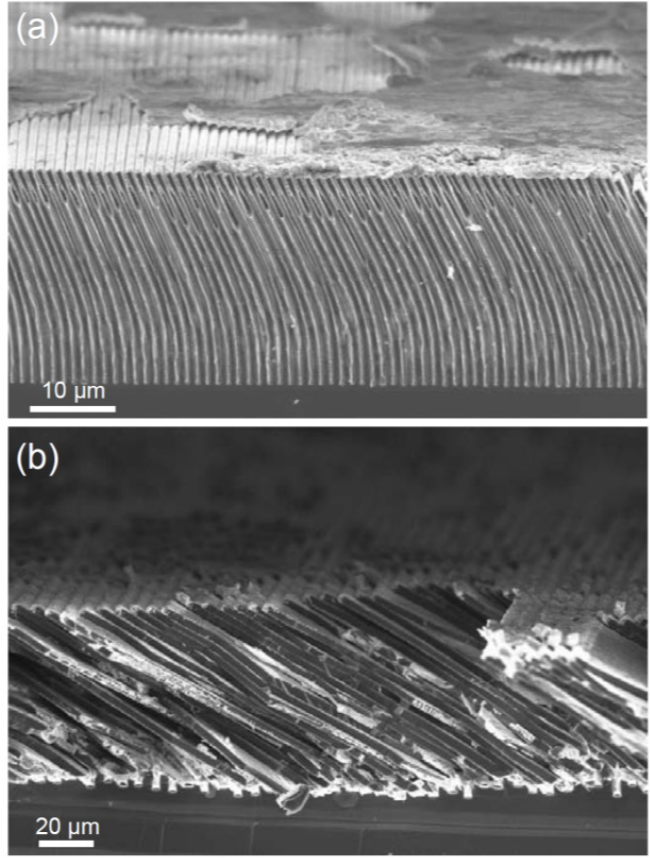

FIG. 8. (a) Grating with pitch of 1 μm and trench depth of 30 μm realized by Bosch process and coated with Ir layer by ALD, the casting was realized with Pb-In at a pressure of 5 MPa; (b) Grating with pitch of 6 μm - height of 140 μm - realized by MacEtch and casted with Pb-In at 5 MPa. Both gratings have aspect ratio of 50:1, indicating the limit for applied pressure on high aspect ratio Si lamellas.

The issue is also related to the stability of the high aspect ratio Si lamellas and it can be solved by introducing some stabilizing structures in the grating geometry. Figure 9 (a) shows the Si grating with pitch size 2.7 μm and trench depth of 40 μm realized by Bosch process, the bridge lines perpendicular to the main grating pattern are indicated by the arrows. The bridge lines have the role to stabilize the Si lamellas during the hot embossing process against shearing perpendicular to the lamellas. Figure 9 (b) reports the 2.7 µm pitch grating filled with Au-Sn by using the embossing pressure of 5 MPa. The method is successfully applied to realize also Pb-In grating, the grating with pitch size of 1 µm and height of 30 µm is reported in Figure 9 (c-d). However, since the bridge lines are limiting the metal flow in the direction of the grating lines, more empty cavities at the bottom of the grating were observed (Fig.9d) indicating that the uniformity of the metal filling can be affected by the presence of the stabilizing structures.

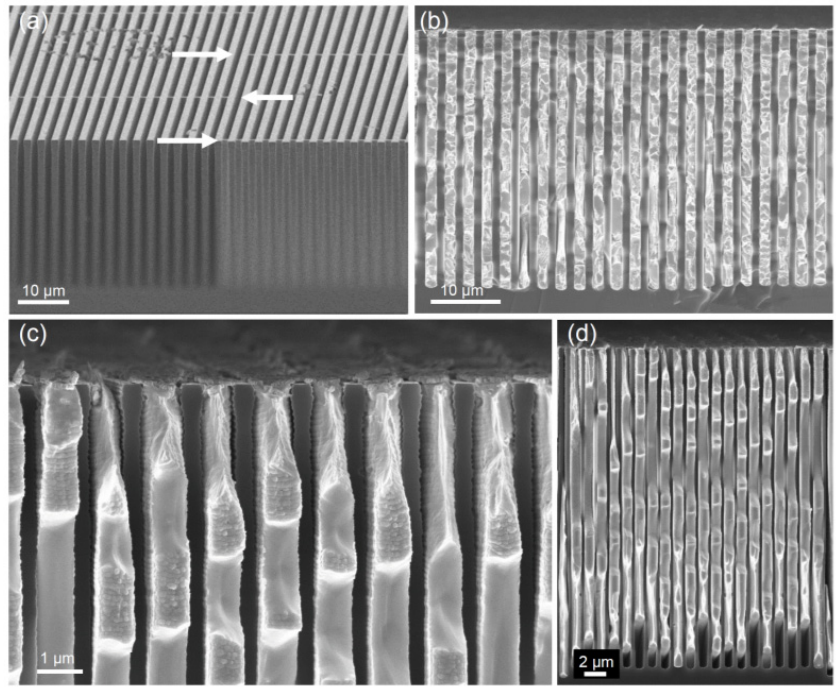

FIG. 9. Si grating (a) with pitch size 2.7 µm and trench depth of 40 µm realized by Bosch process, the bridge lines perpendicular to the main grating pattern are indicated by the arrows; grating filled with Au-Sn (b) by using the embossing pressure of 5 MPa; grating with pitch size of 1 µm and height of 30 µm, hot embossed with Pb-In (c-d) at 5 MPa.

The main results of casting Au- and Pb- based alloys by hot embossing in Si gratings in comparison to conventional Au electroplating of gratings produced by deep X-ray lithography[3] are summarized in the following table.

TABLE I. Summarizing the main metal properties and comparison of processing. The attenuation length is the depth into the material measured along the surface normal where the intensity of x-rays falls to 1/e of its value at the surface. It was calculated using ref.[16] It is assumed that Au electroplating produces pure Au filling of the grating structures, which mean with the same density of bulk Au.

| Metal/alloy | Au electroplating | Au-Sn casting | Pb-In casting | Pb-Sn-Ag casting |
|---|---|---|---|---|
| Composition w% | pure Au | 80 w% Au 20 w% Sn eutectic | 95 w% Pb 5 w% In | 92.5 w% Pb 5 w% Sn 2.5 w% Ag |
| Composition at% | pure Au | 71 at% Au 29 at% Sn | 91.3 at% Pb 8.7 at% In | 87.2 at% Pb 8.2 at% Sn 4.6 at% Ag |
| T [°C] | | *Melting* 280 | *solidus* 292 *liquidus* 314 | *solidus* 287 *liquidus* 296 |
| Density [g/cm$^3$] | 19.32 (bulk material) | 14.32 | 11.06 | 11.02 |
| Attenuation length (X-ray 30 keV) [µm] | 19.98 | 24.29 | 30.98 | 30.81 |
| Attenuation length % | 100 | 121 | 155 | 154 |
| Process | 1. growth of gold within grooves from conducting seed layer; | 1. heating to 280°C and pressure 2. slow cooling (to avoid thermal shock) | 1. heating to 292 °C and low pressure 2. heating to 314 °C high pressure and mold filling | 1. heating to 287 °C and low pressure 2. heating to 296 °C high pressure and mold filling |
| Remarks | polymer instability issues; process time from several hours to several days | needs wetting layer (Ir by ALD); casting process time 30 min – 4h | pressure can already be applied at stage 1; formation of wetting layer at stage 1; casting process time 30 min – 1h | needs wetting layer (Ir by ALD); pressure can already be applied at stage 1; formation of melt with microparticles at stage 1 (possible damage); casting process time 30 min – 1h |
| Aspect ratio (4.8 µm pitch) | 91:1 (ref.[3], with bridges) | 40:1 (ref.[7]) 50:1 (with bridges) | 30:1 50:1 (with bridges) | 20:1 |

For comparison, the attenuation length of Si for X-ray of 20-30 keV is in the range of 1000-3000 μm, so that the absorption from the Si substrate (500 μm thick) of the gratings is negligible for this energy range. However, the Si template can be etched away by using an alkaline solution, this could be useful to produce a full metal grating for the applications where the silicon substrate is undesired (low energy X-ray, bendable gratings etc).

## D. X-ray performance characterization

The use of the gratings in GI experiments will be reported separately. Here, the focus is on the viability of the fabrication and the possible impact of inhomogeneous casting that cannot be detected in SEM characterization. Preliminary characterization of the Au-Sn gratings shows that the alloys behave like expected, with a smaller absorption resulting from the composition of the alloy, i.e. 80 w% for Au-Sn in comparison to electroplated Au gratings.

Due to the X-ray source characteristics, X-ray grating interferometry systems typically consist of a combination of two[1] (synchrotron-based setups) or three X-ray gratings[17] (laboratory-based systems). In the latter case, the system consists of a source absorbing grating (G0), a phase grating (G1), and an analyzer absorbing grating (G2), thus being the casted gratings suitable to be used as absorbing G0 or G2.

In particular, the X-ray performance of Au-Sn casted grating with pitch of 4.8 μm (casting thickness 30 μm, grating area 70×70 mm$^2$) was investigated by a symmetric grating interferometer geometry[17] for a design photon energy of 20 keV at the 3$^{rd}$ Talbot order. The system was built by mounting a set of 3 gratings on a high precision stage.

The total length of the system was 55.8 cm. Initially, the Au-Sn casted grating was used as G0 grating together with a G1 phase grating made of silicon and a G2 absorbing grating fabricated by Au electroplating. An average X-ray fringe visibility of 14.0 %, which is comparable to the values achieved in absorbing gratings fabricated by conventional gold electroplating, was observed. After that, the gratings were exchanged and the Au-Sn casted grating was used as G2, an average X-ray fringe visibility of 16.4 % was obtained. The difference in values is caused by different x-ray filtering properties of the gratings, which suggests that different energy range of x-rays is incident on G1 phase grating, which affects the contrast of interference produced by the gratings. The visibility map (visibility at each X-ray camera pixel) and differential phase contrast and scattering images of sample consisting of polystyrene microspheres of 600 μm are shown in Figure 10. The good quality of the Au-Sn casted grating is observed in the obtained images of the sample.

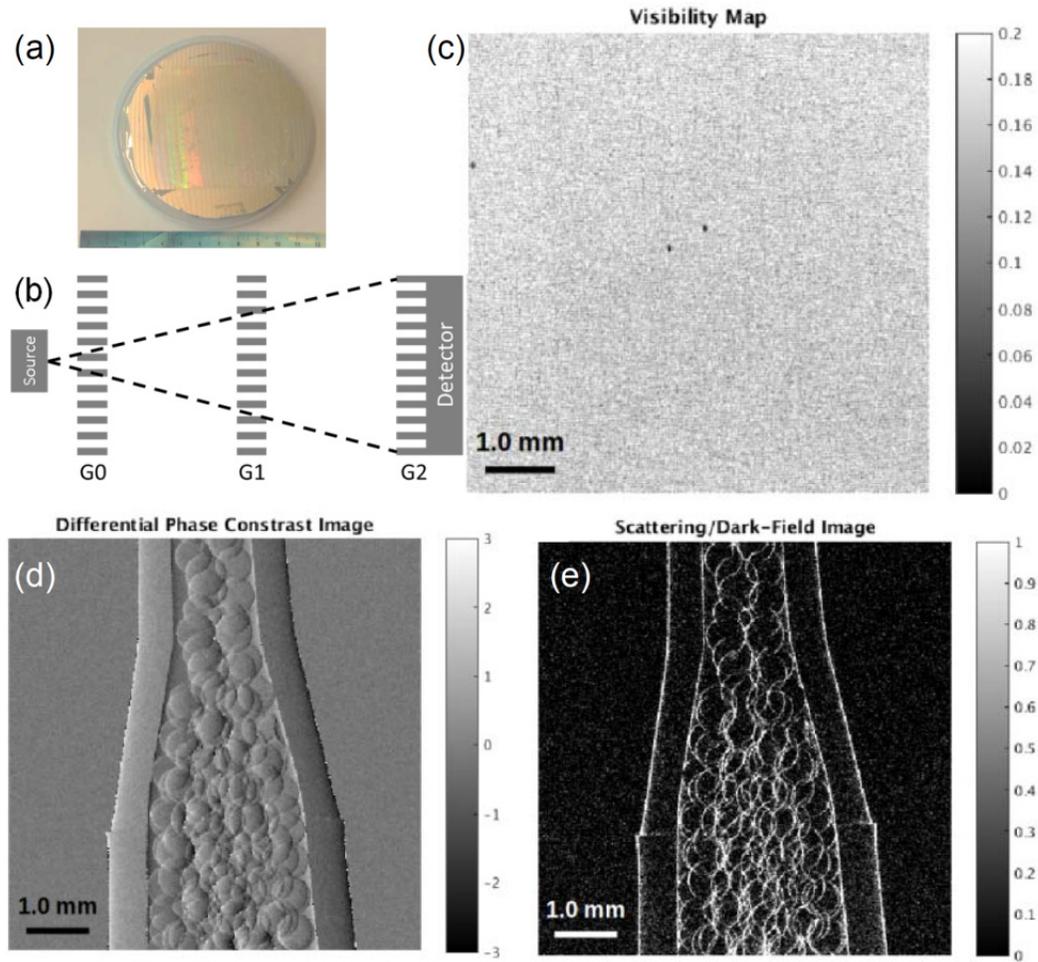

FIG. 10. (Color online) X-ray performance of Au-Sn casted grating of 4.8 µm period used as G2 absorbing grating. Picture (a) of the casted grating (100 mm wafer with a 70x70 mm$^2$ grating area); schematic (not in scale) of X-ray interferometer [17](b); X-ray fringe visibility map (c); differential phase contrast (d) and scattering (e) images of 600 µm polystyrene spheres.

The same kind of characterization has been performed with Pb-In grating, the preliminary results with Pb-In grating (pitch of 4.8 µm, thickness of 50 µm and grating size of 20x20 mm$^2$) as G2 indicate a mean visibility of 12.7 % for a design photon energy of 20 keV at the 3$^{rd}$ Talbot order. Further experiments are necessary to optimize the GI system set up.

## IV. SUMMARY AND CONCLUSIONS

High aspect ratio metal microgratings can be fabricated by direct hot embossing of metal foils into Si gratings. The material properties such as the metal wetting on Si template, the alloy phase diagram, the metal shrinkage during the solidification and the associated stress release contribute to the uniformity and the quality of the manufacturing. The hot embossing parameters such as the heating/cooling ramp and the applied pressure have to be finely tuned as a function of the used metal alloy and grating aspect ratio. We tested Au-Sn, Pb-In and Pb-Sn-Ag alloys underling the different relevant peculiarities. A slow cooling is necessary in order to avoid the Si cracking in Au-Sn casted gratings. This issue is not relevant for Pb-based alloys, being Pb less stiff than Au alloy. Pb-In alloy has a good wettability of the Si surface because of the presence of In that can create a self-wetting layer for the liquid alloy without the necessity of the Ir wetting layer used for the Au-Sn alloy. Pb-Sn-Ag alloy form a liquid-phase containing micro-sized solid particles that damage the Si grating lines during the hot embossing so it cannot be used for grating line size smaller than 3 µm.

The casted gratings were tested in an X-ray GI set up in order to compare their performance with respect to Au electroplated gratings. A visibility in the range of 15 % at 20 keV was obtained using Au-Sn casted gratings. The metal gratings are key components for X-ray phase sensitive GI systems. The new method has relevant advantages, such as being a low cost technique, fast and easily scalable to large area fabrication. The presented low cost and high yield fabrication process has a direct impact to the commercialization of X-ray GI for medical diagnostics and non-destructive testing.

## ACKNOWLEDGMENTS

This work has been partially funded by the ERC-2012-STG 310005-PhaseX grant, ERC-PoC-2016 727246-MAGIC grant. We would like to thank for their valuable collaboration and contributions: K. Vogelsang, S. Stutz, V. Guzenko and C. David from PSI-LMN; M. Kagias, C. Arboleda and Z. Wang from PSI-TOMCAT.

TABLE I. Summarizing the main metal properties and comparison of processing. The attenuation length is the depth into the material measured along the surface normal where the intensity of x-rays falls to 1/e of its value at the surface. It was calculated using ref.[16] It is assumed that Au electroplating produces pure Au filling of the grating structures, which mean with the same density of bulk Au.

| Metal/alloy | Au electroplating | Au-Sn casting | Pb-In casting | Pb-Sn-Ag casting |
|---|---|---|---|---|
| Composition w% | pure Au | 80 w% Au 20 w% Sn eutectic | 95 w% Pb 5 w% In | 92.5 w% Pb 5 w% Sn 2.5 w% Ag |
| Composition at% | pure Au | 71 at% Au 29 at% Sn | 91.3at% Pb 8.7at%In | 87.2at% Pb 8.2 at% Sn 4.6 at% Ag |
| T [°C] | | *Melting* 280 | *solidus* 292 *liquidus* 314 | *solidus* 287 *liquidus* 296 |
| Density [g/cm$^3$] | 19.32 (bulk material) | 14.32 | 11.06 | 11.02 |
| Attenuation length (X-ray 30 keV) [μm] | 19.98 | 24.29 | 30.98 | 30.81 |
| Attenuation length % | 100 | 121 | 155 | 154 |
| Process | 1. growth of gold within grooves from conducting seed layer; | 1. heating to 280°C and pressure 2. slow cooling (to avoid thermal shock) | 1. heating to 292 °C and low pressure 2. heating to 314 °C high pressure and mold filling | 1. heating to 287 °C and low pressure 2. heating to 296 °C high pressure and mold filling |
| Remarks | polymer instability issues; process time from several hours to several days | needs wetting layer (Ir by ALD); casting process time 30 min – 4h | pressure can already be applied at stage 1; formation of wetting layer at stage 1; casting process time 30 min – 1h | needs wetting layer (Ir by ALD); pressure can already be applied at stage 1; formation of melt with microparticles at stage 1 (possible damage); casting process time 30 min – 1h |
| Aspect ratio (4.8 μm pitch) | 91:1 (ref.[3], with bridges) | 40:1 (ref.[7]) 50:1 (with bridges) | 30:1 50:1 (with bridges) | 20:1 |

**Figure captions**

FIG. 1. (Color online) Schematic of the different stages of the fabrication sequence (not to scale): resist pattern by lithography, etched trenches, trenches/grooves with metal conformal coating and completely filled trenches with zero excess material. The top view shows the grating area (70x70 mm$^2$) on a silicon wafer with diameter of 100 mm.

FIG. 2. Pb-In alloy can be separated during melting by a slow heating (10°C/min). The micrographs are high magnification SEM cross sections of grating trenches where: (a) the In-rich nanoparticles are formed on the HF-cleaned Si surface; (b) the liquid Pb does not wet Si as indicated by the contact angle (>90°); (c) the liquid Pb-alloy flows on the segregated In coating, which serves as a wetting layer for the Pb. Therefore, the system can realize a wetting contact without significant variation of the composition or even depletion of the alloy. The grating has pitch of 4.8 µm and was realized by Bosch etch.

FIG. 3. (Color online) (a) Plot of the hot plate temperature, the bottom plate is in contact with the Si template, the top one with the metal foil as indicated in the insert, the applied force is also reported, the pressure is about 5 MPa; (b) Pb-In casted grating, the grating area was 20 × 20 mm$^2$, the pitch is 6 µm and the depth 60 µm, the grating was realized by MacEtch.

FIG. 4. Pb-Sn-Ag alloy (92.5 w% Pb – 5 w% Sn – 2.5 w% Ag) casted in Si grating with pitch size of 4.8 µm realized by Bosch etch and conformally coated by 30 nm Ir. Hot embossing pressure 5 MPa at a maximum temperature of 330 °C.

FIG. 5. Au-Sn (a) casted grating with pitch size of 20 µm; Pb-In (b) casted grating with pitch size of 4.8 µm. In both gratings the large grooves are formed on top of the metal lines because of the metal shrinkage during the cooling down.

FIG. 6. (Color online) Temperature (a) vs time used for fast and slow cooling experiments. Au-Sn casted grating (b-c) obtained with fast cooling rate. The direction of the Si planes (111) and (110) are plotted as a reference in (b), the arrows in (c) indicate the cracking of Si substrate along the (111) and (100) directions and grating pulling off due to the release of the thermal stress. Au-Sn casted grating (d-e) obtained with the slow

cooling rate, the arrow points to delamination, the low magnification image (e) shows no cracks in Si but only delamination.

FIG. 7. Pb-In casting realized with embossing pressure of (a) 0.4 MPa and (b) 1.5 MPa into Si grating with pitch size of 4.8 μm.

FIG. 8. (a) Grating with pitch of 1 μm and trench depth of 30 μm realized by Bosch process and coated with Ir layer by ALD, the casting was realized with Pb-In at a pressure of 5 MPa; (b) Grating with pitch of 6 μm - height of 140 μm - realized by MacEtch and casted with Pb-In at 5 MPa. Both gratings have aspect ratio of 50:1, indicating the limit for applied pressure on high aspect ratio Si lamellas.

FIG. 9. Si grating (a) with pitch size 2.7 μm and trench depth of 40 μm realized by Bosch process, the bridge lines perpendicular to the main grating pattern are indicated by the arrows; grating filled with Au-Sn (b) by using the embossing pressure of 5 MPa; grating with pitch size of 1 μm and height of 30 μm, hot embossed with Pb-In (c-d) at 5 MPa.

FIG. 10. (Color online) X-ray performance of Au-Sn casted grating of 4.8 μm period used as G2 absorbing grating. Picture (a) of the casted grating (100 mm wafer with a 70x70 mm$^2$ grating area); schematic (not in scale) of X-ray interferometer [17](b); X-ray fringe visibility map (c); differential phase contrast (d) and scattering (e) images of 600 μm polystyrene spheres.